\documentclass[%
]{epl}
\usepackage{epsf}
\usepackage[usenames]{color}
\usepackage{amssymb}
\usepackage{amsfonts}
\usepackage{amsmath}

\def\etal{\textit{et al.}}
\def\ie{\textit{i.e.}}

\def\eg{\textit{e.g.}}

\newcommand{\lab} {\left\langle}
\newcommand{\rab} {\right\rangle}

\title{Vibrational effects in the linear conductance of carbon nanotubes}
\shorttitle{Vibrational effects in carbon nanotubes}
\author{M.~Gheorghe\inst{1} \and R.~Guti{\'e}rrez\inst{1} \and N.
Ranjan\inst{2} \and A.~Pecchia\inst{3} \and A.~Di~Carlo\inst{3} \and G.~Cuniberti\inst{1}}
\institute{\inst{1} Institute for Theoretical Physics, University of Regensburg, D-93040
Regensburg, Germany \\ \inst{2} Institute for Physical Chemistry, Technical University of Dresden,
D-01069  Dresden, Germany \\ \inst{3} INFM and Dept.~of Electrical Engineering, University of Rome ``Tor Vergata",
I-00133 Rome, Italy }
\shortauthor{M.~Gheorghe \etal}
\pacs{73.63.-b} {Electronic transport in nanoscale materials and structures}
\pacs{73.63.Fg} {Nanotubes}
\pacs{63.22.+m} {Phonons or vibrational states in low-dimensional structures and nanoscale materials}

\begin{document}

\maketitle

\begin{abstract}
We study the influence of structural lattice fluctuations on
the elastic electron transport in single-wall carbon nanotubes  within a
density-functional-based scheme. In the linear response regime, 
 the linear conductance is calculated via
 configurational averages over the distorted lattice. 
Results obtained
from a frozen-phonon approach as well as
from  molecular dynamics simulations are compared. 
We further suggest that the  effect of
structural fluctuations can be qualitatively captured by the
Anderson model with bond disorder.
The influence of
individual vibrational modes on the electronic transport
is discussed as well as the role of zero-point fluctuations.  
\end{abstract}

\section{Introduction}

Carbon nanotubes (CNTs) have become a paradigm in the physics
of low-dimensional systems due to their fascinating properties~\cite{reich04}.
Especially, the close
interconnection between their chirality and their
electronic structure
 make them an ideal candidate for
applications in the  field of molecular
electronics.
As a consequence, extensive experimental and
theoretical research has been carried out in the past
years to clarify their structural and conducting properties~\cite{reich04}.

Concerning quantum transport in CNTs, it is theoretically well-established that
the linear conductance is quantized in units of $G_0=e^2/h$~\cite{reich04}.
In the case of metallic tubes, effective
low-energy theories as well as
tight-binding and {\it ab initio} calculations have
demonstrated that two massless electronic bands with
linear dispersion cross the Fermi points at ${\bf K}({\bf K^{'}})=
+(-)2\pi /3a_{{\rm 0}}$, $a_{\rm 0}$ being 
the CNT lattice constant~\cite{kane97,reich04}.
As a result, two transport channels per spin are open at the Fermi
level $E_\mathrm{F}$, leading to a conductance of 4$\times
G_0$.
This value is conserved even in the presence of
disorder as far as the range of the impurity potential is larger 
than the nanotube lattice constant~\cite{mceuen99}. 
 The
same is expected to hold in the presence of vibrations at
low but nonzero temperatures, where only long
wavelength modes can be excited~\cite{ando02}. 
This 
is however not  transferable to high temperatures, where additional modes may be
activated, or to higher-lying  bands, where mixing can lead to  
additional backscattering.

So far, the
interaction of electrons with phonon modes in
CNTs has been mainly addressed in periodic
systems ~\cite{figge02}. 
It is rather
difficult to clarify  the interplay between charge transport and 
coupling to vibrational degrees of freedom  in its full generality. This
requires a reliable electronic structure method, the
calculation of  electron-phonon matrix elements and the
combination with a transport formalism.
Some developments in this direction have recently been
presented~\cite{emberly00},
though the calculations were limited to small
molecular systems.
In nanotubes, however, the large
number of vibrational modes make these approaches
computationally very  demanding.

Here, we investigate  the influence of structural  fluctuations  
on the {\it elastic} electron transport 
in single-wall  CNTs. We apply 
a recently proposed computational scheme \cite{pecchia03} 
to shed some light on the influence of vibrational modes on charge propagation in
CNTs.
The method combines  Green function techniques with a density-functional-based 
(DF) methodology
for the electronic  properties of the
system. Electron-phonon matrix elements are not 
directly calculated; the effect of the lattice distortions on charge 
propagation is considered via suitable configurational averaging procedures. 

Anticipating our main results, 
 we find that the global effect
of thermal fluctuations turns out to be  stronger for massive bands
than for massless bands, \ie~the transmission
around the Fermi energy is not appreciably affected by
them on the length scales investigated here.
Further, we show that  the
effect of the atomic vibrations on the electron transport can
be qualitatively captured by static disorder in the spirit of 
the Anderson model~\cite{anantram98}.

\section{System and Methodology}
\begin{figure}[t]
\centerline{
\includegraphics[width=2.4in,clip,angle=0]{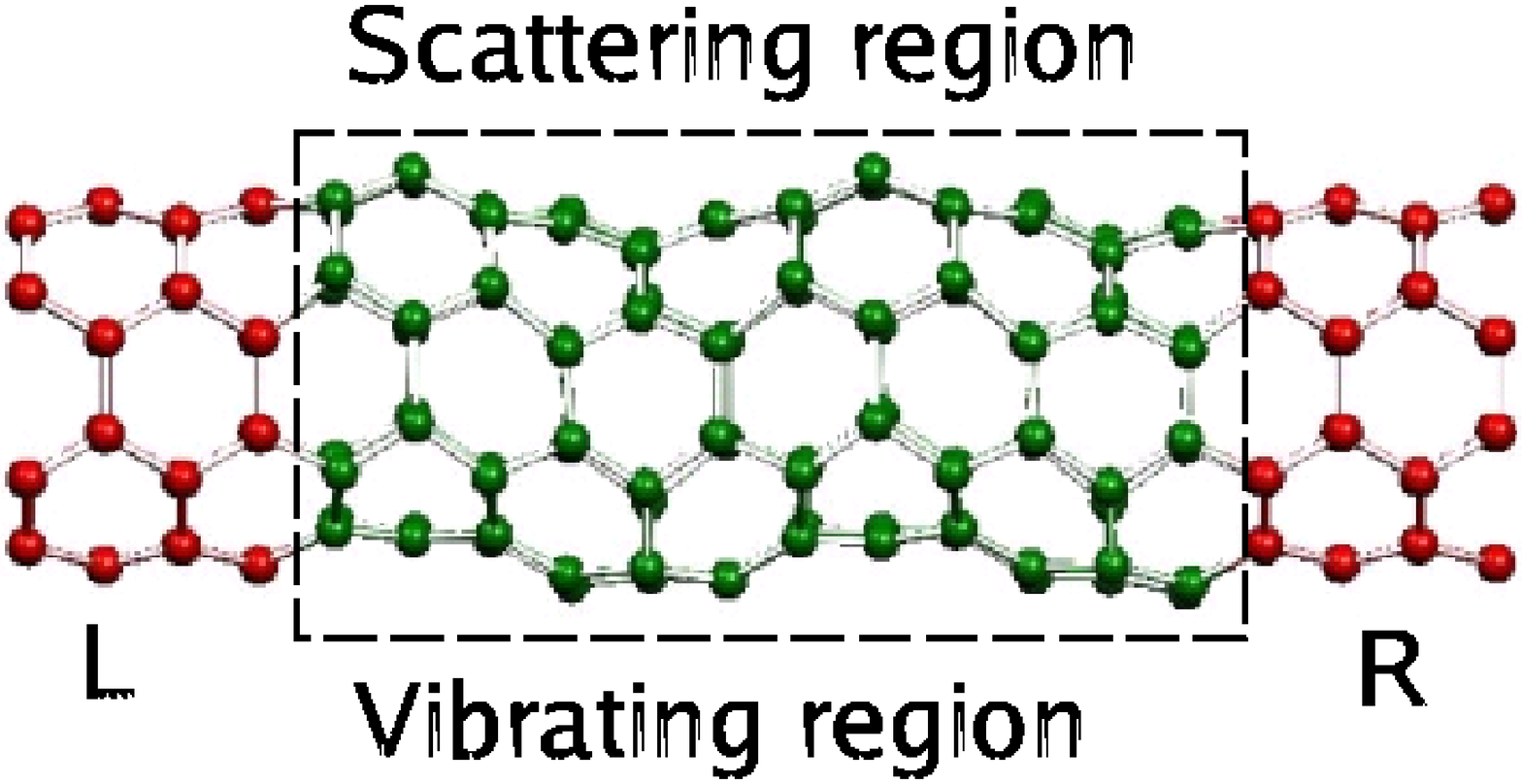}
}
\vspace{0.35in}
\centerline{
\includegraphics[width=2.75in]{./fig1b.eps}\hspace{0.25cm}
\includegraphics[width=2.75in]{./fig1c.eps}
}
\caption{\label{fig1} Top: snapshot of
the vibrating part (scattering region (green)) of an infinite metallic (4x4) CNT. 
The semiinfinite left (L) and right (R) segments act as electrodes. 
Bottom:  temperature dependence of the conductance 
spectrum for a (4,4) CNT
(a) and corresponding
DOS (b), in the
frozen-phonon approach. 
Panels (c) and (d) display the results from   molecular dynamics 
 simulations for the same geometry. ``Ground state'' refers to the zero
 temperature configuration.
}
\end{figure}

The system  consists of an
infinite metallic (4,4) CNT, where a finite section is allowed
to vibrate and thus define the {\em scattering region}, see Fig.~\ref{fig1}. 
The remaining semiinfinite
segments of the nanotube constitute the electrodes. This configuration 
approximately mimics the experimental situation of clamped CNTs
\cite{babic03}.
An advantage
 of this geometry is the
 possibility of comparing the conductance in presence of thermal
  fluctuations
 with the limiting case of a perfect infinite tube, 
 where conductance quantization
  is  
 obtained. 
 Calculations  for semiconducting  tubes were also
performed \cite{ggc04} and showed that the gap region 
is not appreciably affected by vibrations, while outside  
the gap a behavior quite 
similar to that 
presented here is obtained. Because of the large number 
of atoms required to
simulate the electrodes, the interface and the vibrating region (scattering
region), 
only short sections of the nanotube were
included in the latter (three unit cells). Numerical tests with 
up to six unit cells 
 have  not shown 
any new qualitative effects when comparing with the present calculations. 
With
increasing number of  cells  the phonon spectrum will
 develop features of the
infinite system, \eg~emergence of precursors of 
low-frequency acoustic and optical modes. 
Obviously, our approach can not well describe  these long-wave length modes; the main 
effects we find here are however related to vibrations with energies larger than
$80-100$ meV 
($\sim 640-800 \, {\rm cm}^{-1}$).

%%%%%%%%%%%%%%%%%%%%%%%%%%fig1%%%%%%%%%%%%%%%%%%%%%%%%%%%%%%%%%%%%%%%%%%%%%%%%
%%%%%%%%%%%%%%%%%%%%%%%%%%%%%%%%%%%%%%%%%%%%%%%%%%%%%%%%%%%%%%%%%%%%%%%%%% 

The equilibrium geometry
 at zero temperature was found by conjugate
gradient relaxation techniques with a
DF-parametrized tight-binding 
Hamiltonian~\cite{thomas00}. For a carbon-based system, we use a minimal $2s2p^{3}$
valence  basis set to expand the electronic eigenstates. 
All energies are
measured henceforth with respect to the Fermi level. 

 The vibrational degrees of freedom  of the scattering region are
described within the harmonic approximation. We expect
the vibrational modes of CNTs to be
well-represented by harmonic potentials, in contrast
to previously studied organic molecules~\cite{pecchia03}, where low-frequency
anharmonic modes were present (torsional modes).

%%%%%%%%%%%%%%%%%
  Our approach accounts for the
influence of  structural fluctuations on the charge transport by 
 averaging over sets of atomic configurations $\{\delta {\bf{r}}_{\ell}\},
\ell=1,\cdots N$, where $N$ is the number of atoms in the scattering region.
We can thus define  a configurational  averaged {\it elastic} linear conductance 
 $ g(E)=2G_0 \lab T(E,{\delta \bf{r}_{\ell}})\rab$~\cite{pecchia03}.
Given an atomic   configuration $\{\delta \bf{r}_{\ell}\}$, the 
transmission  
 can be calculated as
$T(E,{\delta \bf{r}_{\ell}})=
\mathrm{Tr}\,\{G^{\dagger}({\delta \bf{r}_{\ell}})\, \Gamma_\textrm{R} \,
G({\delta \bf{r}_{\ell}})\, \Gamma_{\textrm L}\}$~\cite{cgg02}.
The Green function of the scattering region is
given by
$
G^{-1}(E,\delta {\bf{r}}_{\ell})=E S({\delta {\bf{r}}_{\ell}}) -H({\delta
{\bf{r}}_{\ell}})-\Sigma_{\textrm{L}}-\Sigma_{\textrm{R}}.
$
 $H$ and $S$ are the Hamiltonian and overlap matrices
of the vibrating region with configuration $\{\delta \bf{r}_{\ell}\}$.
 The overlap
matrix takes into account the non-orthogonality of the used basis set. 
Notice that the influence of the  leads 
has been transferred  now into complex self-energy functions
$\Sigma_{\textrm{L,R}}$.
 Finally, 
$\Gamma_{\textrm{L,R}}=i(\Sigma_{\textrm{L,R}}-\Sigma^{\dagger}_{\textrm{L,R}})$ 
are the corresponding spectral
densities  of the leads \cite{cgg02}. 

The linear conductance is calculated within two complementary
schemes based on a quantum mechanical resp. classical treatment of the
vibrational modes~\cite{pecchia03}:
(i) a frozen-phonon approach (FPA) and (ii) and  DF-based  molecular dynamics
(MD) simulations. Especifically, in the
FPA  the scattering region is statically distorted according to the eigenvectors 
of the phonon
modes obtained by diagonalizing  the dynamical matrix, i.~e. via the expansion 
$\delta {\bf{r}}_{\ell} =\sum_{\alpha=1}^{3N} x_{\alpha}
{\bf{e}}^{\alpha}_{\ell}$ 
with 
$\bf{e}^{\alpha}_{\ell}$ being the mode eigenvectors. The $x_{\alpha}$
chracterize the amplitude of
the atomic displacements, distributed according to 
$P(\{ x_{\alpha} \})=\prod_{\alpha}
(\sigma/\sqrt{\pi}) \exp(-\sigma^2\,x_{\alpha}^2)$, with 
 $\sigma^2=m_{\alpha} \omega_{\alpha}^2/2E_{\alpha}(T)$ and 
$E_{\alpha}(T)=\hbar\omega_{\alpha}(N_{\alpha}(T)+1/2)$ 
being  the  energy of a normal
mode~\cite{pecchia03}. 
Note that the temperature dependence enters via the Bose factors
$N_{\alpha}(T)$.
Using a Monte-Carlo sampling
technique, an average conductance $g_{{\rm FP}}(E)$
is calculated.
In the MD approach,  averages
are taken over the simulation time leading to a
$g_{{\rm MD}}(E)$.
  Both approaches assume the adiabatic approximation to hold.
 An estimation of  time scale ratios yields $\tau_{phon}/\tau_{el}(E_{\rm F})\sim 30-60$ 
 for phonon energies $\ge 60$ meV. Very low-frequency modes may become
 problematic, but they are not well
 described when using short CNT sections, as previously mentioned.

%%%%%%%%%%%%%%%%%% CONDUCTANCE %%%%%%%%%%%%%%%%%%%%%%%%%%%
%%%%%%%%%%%%%%%%%%%%%%%FIGURE 2%%%%%%%%%%%%%%%%%%%%%%%%%%%%%%%%%%%%%%%%%%%%%%%%%%%%
\begin{figure}[t]
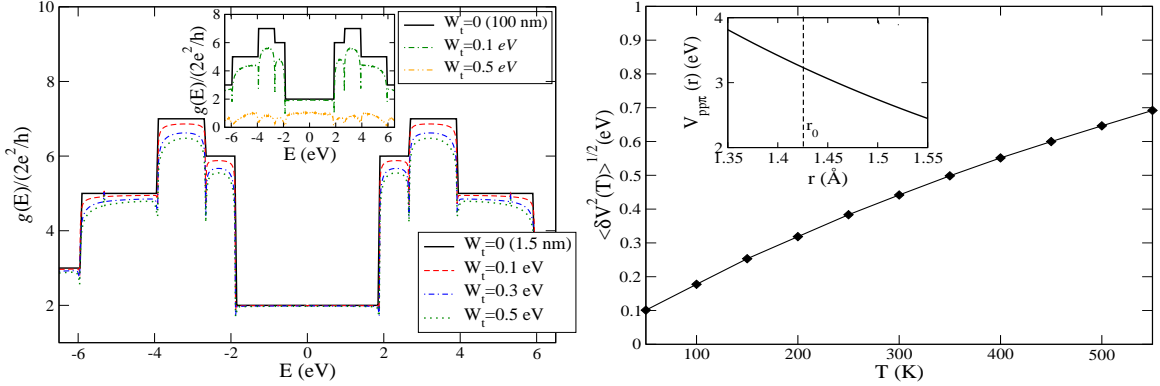

\centerline{
\includegraphics[width=3in,height=2in,clip,angle=0]{./fig2a.eps}
\includegraphics[width=3in,height=2in,clip,angle=0]{./fig2b.eps}
}
\caption{\label{fig2} 
Left panel: average conductance obtained from the Anderson
Hamiltonian with bond disorder for a short (4 nm) (4x4) CNT. 
The inset shows similar results for  a much longer CNT (100 nm). Right panel:
temperature dependence of the average mean-square fluctuation $\sqrt{\lab
V^{2}\rab(T)}$ of the $\pi-\pi$ hoppping
integral. The inset shows the corresponding distance dependence of this matrix
element.
} 
\end{figure}
%%%%%%%%%%%%%%%%%%%%%%%%%%%%%%%%%%%%%%%%%%%%%%%%%%%%%%%%%%%%%%%%%%%%%%%%%%%%%%%%%%%%%%%

\section{Results}

Fig.~\ref{fig1} shows the average conductance as
a function of the energy of the incoming electron,
calculated for the FP and MD schemes.
Both methods yield qualitative similar results.
For a
CNT at zero temperature and without inclusion of atomic fluctuations, perfect
conductance quantization is found, as evident from Figs.~1(a)~and~1(c)
(solid lines). 
The corresponding DOS displays the typical 
van-Hove singularities, see Figs.~\ref{fig1}(b) and \ref{fig1}(d) (solid lines). 

When taking into account atomic motion, 
the perfect conductance quantization is
washed out, however.
Several features can be seen in
Fig.~\ref{fig1}.
The temperature dependence of $g_{{\rm
FP}}(E)$ is almost negligible on the lowest
conductance plateau around  the Fermi energy, 
indicating that backward scattering is considerably weakened 
 in the low-energy sector of the spectrum.
Higher-lying  bands are however more affected,
the conductance being drastically reduced already at 1 K, see Fig.~\ref{fig1}(a). 
Note, however, that upon this 
initial suppression these bands are not 
very sensitive to further temperature variations. The reason is that 
the main contribution to the conductance arises from 
modes with energies larger than $100$ meV ($\sim 1000 K$),  
whose thermal occupation factors are  much less 
than one for the temperature range discussed here (see below).
At the crossover points, where new bands
start contributing to transport, \eg~at energies
~$-2.48~e$V, $-1.18~e$V, and $1.15~e$V, conductance dips are
found.
The corresponding DOS, Fig.~\ref{fig1}(b),
shows strong broadening  of the van-Hove singularities, implying a shift of the spectral weight
to their neighboring region.
The increased number of states around
the crossover points enlarges the phase space for
backscattering resulting into the previously mentioned 
conductance dips. 

%%%%%%%%%%%%%%%%%%%%%%%FIGURE 3%%%%%%%%%%%%%%%%%%%%%%%%%%%%%%%%%%%%%%%%%%%%%%%%%%%%
\begin{figure}[t]
\centerline{
\includegraphics[width=3in,clip,angle=0]{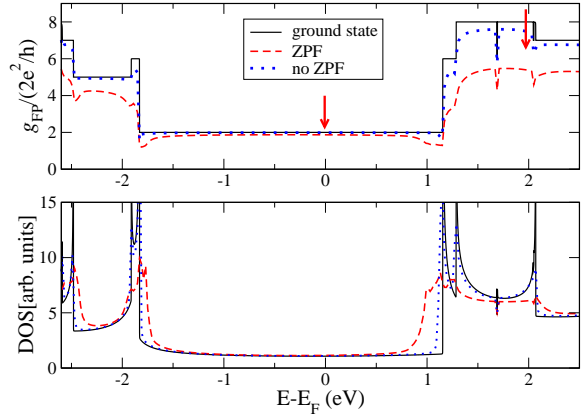}}
\caption{\label{fig3} The influence of zero-point fluctuations on the conductance 
of the metallic (4,4) nanotube at a fixed temperature ($500$ K). Note the strong 
broadening of the Van-Hove singularities which causes strong conductance dips. 
}
\end{figure}
%%%%%%%%%%%%%%%%%%%%%%%%%%%%%%%%%%%%%%%%%%%%%%%%%%%%%%%%%%%%%%%%%%%%%%%%%%%%%%%%%%%%%%%
 
Let us  now consider  the  results from MD
 simulations, see  Fig.~\ref{fig1}(c) and (d).
Though the basic
features found in the FPA  are also seen here, there is
however a less abrupt suppression of the conductance
of the massive bands with increasing temperature.
Thus, at $T=1 \,K$ basically no differences to the zero-temperature case can be 
seen. Even at  $500\, K$,  $g_{\rm MD}$ is much closer to the
zero temperature case, in contrast to the FPA.
This is related to the zero-point
fluctuations (ZPF), which are included in the quantum 
mechanical FP calculation,  but are
absent in the classical MD approach.  The zero-point energy ($\sim
\hbar\omega/2$) is inversely
proportional to the square root of the atomic mass, so that we may expect that 
the impact of ZPF on the conductance for a light atom like carbon will be rather strong.
Nonetheless, the similarity of the
results for the FP and MD approaches indicates that
the harmonic approximation is reliable when dealing
with the vibrational spectrum of CNTs.

%% ANDERSON %%%%%%%%%%%%%%%%%%%%%%%%%%%%%%%%%%%%%%%%%%%%%

In our calculations, an electron propagating
 along the vibrating part of the tube will basically feel a quasi-random field. 
We might expect that some relation to well-known models for disorder may exist.
To test this, we  have performed calculations based on the 
Anderson model within a simple $\pi$-orbital 
approximation~\cite{anantram98}.
 The Anderson Hamiltonian  reads :
$ H=\sum_{j} \epsilon_j c^{\dagger}_{j} c_{j} - \sum_{j,l} t_{ lj}[ c^{\dagger}_{j}
 c_{l} + {\rm H.~c.~} ]$,
where the operators $c^{\dagger}_{j}$($ c_{j}$) destroy (create) 
an electron on the
$\pi$-orbital at site $j$.
Considering only the $\pi$-orbital subspace yields reliable results 
for CNTs with not too small radii, where $\sigma-\pi$ hybridization can be
neglected. Though both, onsite and bond
disorder may be considered, we only present results for the latter case. Bond
 disorder is expected to better mimics the physical situation 
 we are considering here,
 \ie~atomic vibrations, which should mainly influence the C-C bond lengths. 
 Onsite disorder gives qualitative similar results; we thus  set $\epsilon_j=0$
 in what follows. 
The hopping integrals were  randomly drawn 
from  the interval $[ -W_t+t_{\rm hop},W_t+t_{\rm hop} ]$, 
where $t_{\rm hop}\sim 2.66$ eV is a typical hopping
integral for carbon. 

%%%%%%%%%%%%%%%%%%%%%%%FIGURE 4%%%%%%%%%%%%%%%%%%%%%%%%%%%%%%%%%%%%%%%%%%%%%%%%%%%%
\begin{figure}[t]
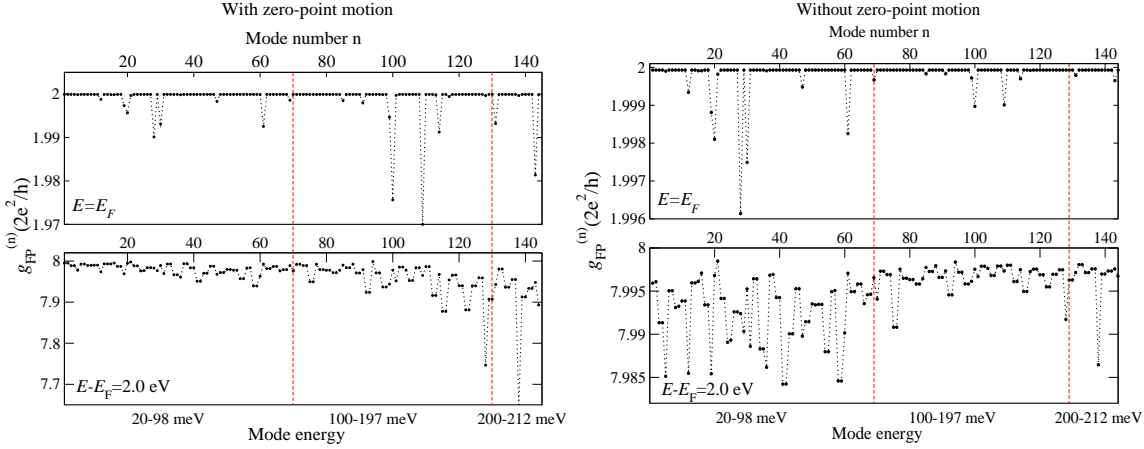

\centerline{
\includegraphics[width=2.95in,height=2.50in,clip,angle=0]{./fig4a.eps}\vspace{0.5cm}
\includegraphics[width=2.97in,height=2.50in,clip,angle=0]{./fig4b.eps}
}
\vspace{-0.8cm}
\caption{\label{fig4} Single mode analysis of the conductance at two selected 
electronic energies, indicated by arrows in Fig.~\ref{fig3}, at $T= 500 \, K$. The
horizontal axis labels the mode number $n$ as well as its energy. Each point 
in the plot corresponds to the linear conductance 
$g_{{\rm FP}}(E)$
%\neq\sum_{n}g^{n}_{{\rm FP}}(E)$
at a given energy $E$
of the incoming electron when just the mode number $n$ is included in the 
calculation and all other modes excluded.
We additionally compare cases with and without inclusion of 
zero-point fluctuations. 
}
\end{figure}
%%%%%%%%%%%%%%%%%%%%%%%%%%%%%%%%%%%%%%%%%%%%%%%%%%%%%%%%%%%%%%%%%%%%%%%%%%%%%%%%%%%%%%%

We have performed calculations for  short (4 $nm$) and long (100 $nm$) nanotubes, see
Fig.~\ref{fig2}, left panel.
 One clearly sees that the two basic signatures
of structural fluctuations previously found, namely 
(i) conductance dips at the band crossover
points and 
(ii)  conductance suppression on the  massive bands, are
 qualitatively reproduced within the Anderson model. 
 The influence of disorder on the conductance 
 is, as expected, dependent on the CNT lenght. 
 This is clearly seen at the charge
 neutrality point, where $g(E_{\rm F})$ is much stronger suppressed for longer tubes. 
 From these results, we may expect a relation between 
 $W_t$ 
and the strength of thermal fluctuations, gauged by $k_{\rm B}T$. This will 
lead to a $T$-dependent disorder parameter $W_{t}(T)$. However, the differences in the basis
sets used in the FPA/MD approaches (non-orthogonal $2s2p^3$ basis) and in 
the  Anderson model ($\pi$ orbitals), respectively,
makes a straightforward mapping very difficult. 
We may get some insight by estimating the temperature dependence
of  the $V_{pp\pi}$ matrix elements in our DF-based method 
\cite{thomas00}. The $V_{pp\pi}$ integrals basically describe the interaction
 of the
$p_z$ orbitals and can be thus related to the empirical $\pi$-orbital models. 
 We  expand  them to linear order in the atomic displacements
around the equilibrium C$-$C bond, $r^{0}_{\rm CNT}\sim 1.43\, \AA$,  
for each atom $\ell$ in the scattering region: 
$\delta\, V_{\ell}=V_{pp\pi}({\bf r}_{\ell})-V_{pp\pi}(r_0)\approx(dV_{pp\pi}/d{\bf r})_{0}\, 
\delta {\bf r}_{\ell}$. 
From the Monte-Carlo sampling outlined above~\cite{pecchia03}, 
it is possible to compute an average
 mean-square fluctuation of the atomic distortions
$\lab \delta {\bf r}^2 \rab(T)=\sum_{\ell=1}^{N}\lab \delta {\bf r}_{\ell}^2
\rab(T)$, where $\lab\cdots\rab$ is an average over the $P(\{ x_{\alpha} \})$ 
distribution (see above) and a further average over all atoms in the scattering region 
has been carried out. 
The related T-dependent fluctuations $\sqrt{\lab \delta\,
V^{2} \rab(T)}$,
 give a measure of the degree of bond disorder introduced by the thermal
motion and are thus related to the Anderson parameter $W_{t}$,
see the right panel of Fig.\ref{fig2} .

%%%%%%%%%%%%%%%%%%% ZPF %%%%%%%%%%%%%%%%%%%%%%%%%%%

Next, we address in more detail the influence of ZPF on the conductance. In
Fig.~\ref{fig3} we show the averaged conductance for a fixed temperature 
with and without ZPF. We clearly see that ZPF do not have a sensitive influence 
on the behavior around 
the Fermi energy, i.~e.~on the lowest conductance plateau.
However, the neglect of ZPF  appreciably weakens  the conductance 
suppression on the massive  bands. 
%%%%%%%%%%%%%% SINGLE MODES %%%%%%%%%%%%%%%%%%%%%

An important advantage of the FPA  is the
possibility to isolate the influence of {\em individual}
vibrational modes on the electronic transport at a
given temperature.
In Fig.~\ref{fig4}, we show the 
conductance $g^{(n)}_{{\rm FP}}$ when {\em only} one eigenmode at the time 
is included in the transport calculation. 
 In the figure, the $x$-axis
labels both, the mode number $n$ and its energy.  
Results are
shown for two different electronic energies, indicated by
arrows in Fig.~\ref{fig3}. A
general feature we can clearly see, is that electrons
at the Fermi energy do not ``see'' the 
vibrational field, \ie~only few modes  give a
contribution to the conductance; moreover, the induced conductance
change  is much smaller than one percent.
This illuminates farther our previous observation that
near $E_{\rm F}$ the structure of the electronic
spectrum is not appreciably changed when compared with
the zero temperature case.
The  other energy showed corresponds to a high-conductance plateau  around  
$2.0~e$V.
On the latter, almost all vibrational modes are
contributing.
The  reason is that enhanced backscattering related to channel mixing is more
effective for massive bands and so the phase space available for scattering
becomes larger. As shown in Fig.~\ref{fig4} (right panel), the neglection of 
ZPF  reduces the contribution of the higher-lying  modes, which are 
mainly C-C stretching bonds with different spatial patterns.
The reason is that at the temperatures considered here (up to $500$ K) the
thermal occupation
 of  these modes is basically negligible. As a result, 
the main contribution to the mode energy  arises from the ZPF terms $\sim
\hbar\omega/2$. 
 Why the global conductance 
suppression {\em without} ZPF becomes much weaker,  can be easely understood by looking at 
  the 
typical length scales of the problem. The wave length of a tunneling electron 
is of the order of $0.5\, nm$, which is approximately of the same order as the 
length of the scattering region $l_{{\rm sc}}\sim 0.7 \, nm$. Low-frequency modes have 
wave lengths  larger than $l_{{\rm sc}}$, the opposite holding for high-energy modes. 
Hence, if the latter are inactive (no ZPF and negligible thermal factors), a propagating charge 
will mainly ``see''  long-wavelength distortions, 
which are  less effective in scattering electrons.

\section{Conclusions}

We have investigated  signatures of structural distortions 
 in the conductance spectrum of a metallic CNT.  
We found that the average effect of the lattice  fluctuations  may be qualitatively described 
 by  Anderson disorder. Our results point out that the linear bands crossing 
the degenerate Fermi points are not appreciably  
affected by structural fluctuations 
within the temperature range and  length scales considered here. As a result, 
the theoretically expected conductance of 4$\times G_0$ is obtained. Massive 
electronic bands are however much more  perturbed, their conductance being strongly 
reduced when comparing with the values of perfect CNTs. 

The authors thanks P. Pavone for fruitful discussions. M. G. thanks the University
of Regensburg for financial support.
This work has been supported by the Volkswagen foundation.

%%%%%%%%%%%%%%%%%%%%%%%%References%%%%%%%%%%%%%%%%%%%%%%%%%%%%%%%%%%%%%%%%

\end{document}